# Detecting and Preventing "Multiple-Account" Cheating in Massive Open Online Courses

Curtis G. Northcutt [1, *], Andrew D. Ho [2], Isaac L. Chuang [1]

[1] *Massachusetts Institute of Technology, Cambridge, MA 02139, USA; {cgn, ichuang}@mit.edu*
[2] *Harvard University, Cambridge, MA 02138, USA; andrew_ho@gse.harvard.edu*

**ABSTRACT**

We describe a cheating strategy enabled by the features of massive open online courses (MOOCs) and detectable by virtue of the sophisticated data systems that MOOCs provide. The strategy, Copying Answers using Multiple Existences Online (CAMEO), involves a user who gathers solutions to assessment questions using a "harvester" account and then submits correct answers using a separate "master" account. We use "clickstream" learner data to detect CAMEO use among 1.9 million course participants in 115 MOOCs from two universities. Using conservative thresholds, we estimate CAMEO prevalence at 1,237 certificates, accounting for 1.3% of the certificates in the 69 MOOCs with CAMEO users. Among earners of 20 or more certificates, 25% have used the CAMEO strategy. CAMEO users are more likely to be young, male, and international than other MOOC certificate earners. We identify preventive strategies that can decrease CAMEO rates and show evidence of their effectiveness in science courses.

## INTRODUCTION AND MOTIVATION

Massive Open Online Courses (MOOCs) began receiving significant media coverage in 2012 (Pappano, 2012; McNutt, 2013), coincident with the widespread commitment by established universities to providing free courses online (Ho et al., 2014; Christensen, Steinmetz, Alcorn, Bennett, Woods, & Emanuel, 2013; Stanford Online, 2013). These MOOCs distinguished themselves from predecessors like MIT's Open Courseware (d'Oliveira, Carson, James, & Lazarus, 2010; Smith, 2009) by providing not only free content but a course-like structure, including enrollment, synchronous participation, periodic graded assessments, online discussion forums, interactive simulations, and of greatest relevance for our purposes, certification of successful completion (DeBoer, Ho, Stump, & Breslow, 2013; Linn, Gerard, Ryoo, McElhanney, Liu, & Rafferty; 2014). One theory of MOOC proliferation holds that free certification of proficiency in college courses can reduce inefficiencies in higher education by replacing high-cost residential courses with low-cost online certification (Hoxby, 2014).

In this paper, we reveal a particular cheating strategy that is prevalent in MOOCs and currently presents a serious threat to the trustworthiness of MOOC certification. We call the strategy, Copying Answers using Multiple Existences Online (CAMEO). A user employing this strategy, whom we refer to as a CAMEO user, earns a certificate by creating at least two MOOC accounts: (1) one or more "harvester" accounts used to acquire correct answers by guessing at test answers and then accessing instructor-provided solutions via a "Show Answer" button, and (2) one or more "master" accounts used to submit these solutions as correct test answers.

---

[*] Please direct correspondence to cgn@mit.edu.

The CAMEO strategy is unique to MOOCs, but lies at the intersection of a number of other copying techniques and contexts. We distinguish between 1) what is copied, 2) why it is copied, 3) how it is copied, and 4) how copying is detected. CAMEO is most similar to "multiple account" sharing strategies in online games (e.g., Kafai & Fields, 2009), where a single user can increase scores or other in-game outcomes by creating multiple accounts and interacting them strategically. However, CAMEO behavior distinguishes itself from online game strategies due to what is copied (correct answers to tests) and why it is copied (to fake or expedite certification of proficiency). As we show, the specificity of these differences enables targeted detection, quantification, and prevention of CAMEO use in MOOCs.

Cheating by CAMEO shares similarity in purpose with copying in online and conventional courses (McCabe, Butterfield, & Treviño, 2012; Palazzo, Lee, Warnakulasooriya, & Pritchard, 2010; Baker, Corbett, & Koedinger, 2004; Mastin, Peszka, & Lilly, 2009; Kauffman & Young, 2015). However, three features of CAMEO make it a unique threat as a cheating strategy in online education. First, it is internally sufficient. Whereas most users copy from other students or external resources, CAMEO users employ multiple accounts to copy from themselves, making the cheating strategy highly accessible by removing dependence on outside resources. As a result, the strategy is extremely effective. Second, in asynchronous MOOCs, where students can access course materials and assessments at their own pace, a CAMEO user can employ the CAMEO strategy for every question they attempt, allowing certification for full course completion in a single sitting. Third, it is unrestricted, employable in a nonselective, open admission setting. Degrees from selective institutions certify, at the very least, that users have been pre-screened, but MOOC certificates do not. Because MOOC users, unlike most postsecondary students, are not selected by any merit-based process or criteria, the considerable accessibility of CAMEO also holds the potential to render the MOOC certificate valueless as an academic credential.

A key contribution of this paper is a detection algorithm for the CAMEO-based cheating that allows for a lower bound estimate of prevalence. This complements the considerable survey literature on cheating, whose estimates may be influenced by social desirability, interpretation of item prompts, and concerns about anonymity. This paper investigates a specific cheating strategy using an algorithm customized to big datasets that record detailed user interactions with online course content, including activity timestamps.

CAMEO also represents an example of a more general tendency for open online learning systems to enable both new strategies for cheating and new strategies for detection. Although CAMEO is a copying strategy, we argue that its use in MOOCs constitutes "cheating." At a minimum, employing CAMEO is a violation of policy, because MOOC honor codes forbid the creation of multiple accounts (Coursera, 2012; edX, 2014; Udacity, 2014). The CAMEO strategy also threatens perceptions of the value of MOOC certification. Any reasonable interpretation of standard MOOC certificates, which refer to "successful completion" (edX, 2015), includes proven student proficiency with course content. Yet, the prevalence of the CAMEO strategy justifies a starkly contrasting interpretation of MOOC certification—that a user merely copied answers from a "dummy" harvester account. Combined with growing evidence that the reputation and usefulness of MOOC certification are predictors of MOOC persistence (e.g., Alraimi, Zo, & Ciganek, 2015), we can conclude that widespread awareness of the MOOC susceptibility to the CAMEO strategy could depress MOOC popularity and persistence among general users.

## METHODOLOGY

The CAMEO detection algorithm relies on the distribution of differences in time between particular user actions across particular user pairs. We first define this difference in time as it relates to CAMEO, then present an algorithm for identifying CAMEO users based on a Bayesian criterion of the timestamp difference distributions. Including the criterion, the CAMEO detection algorithm is comprised of five filters with highly conservative cutoffs intended to reduce false positives.

### *Defining "Copying Answers using Multiple Existences Online" (CAMEO)*

Fig. 1 illustrates two prototypical CAMEO users, each with two accounts, and their timeline of interactions with online assessments. For both CAMEO users in Fig. 1, we also illustrate the variable:

$$\Delta t_{m,h,c,i} = t_{m,c,i} - t_{h,c,i}$$

This is the difference between the time that a master account, $m$, submits a correct answer and the time that a harvester account, $h$, acquires the correct solution, for a problem (item) in common, $i$, in a given MOOC course, $c$. It is possible for a single master to have multiple harvesters and a single harvester to have multiple masters. The subscript, $c$, recognizes that the same master-harvester pair may be employing CAMEO across multiple courses.

Logically, for CAMEO users, these $\Delta t$ are predominantly or entirely positive in sign. The former time, $t_{m,c,i}$, is recorded in server log files. For the latter time, $t_{h,c,i}$, we take advantage of the fact that instructors of the MOOCs in our sample generally allow users to click a "show answer" option after submitting answers, to display a staff-prepared answer and/or an explanation of the solution, in order for users to obtain rapid feedback. The timestamp produced by a "show answer" click defines $t_{h,c,i}$. We introduce a method for probabilistic detection of CAMEO users based on observed distributions of $\Delta t_{m,h,c}$ over items $i$.

### *Detection of "Copying Answers using Multiple Existences Online" (CAMEO)*

The detection strategy begins by considering all possible ordered pairs of accounts, within each course, as candidate CAMEO users. It asks whether the pattern of "show answers" from one, the "candidate harvester" (CH), and "correct answers" from the other, the "candidate master" (CM), is ordered and coincident enough to declare the CH-CM pair a CAMEO user. In total, we employ five filters to identify CAMEO users (Table 1). These five filters are conjunctive and thus order-independent; we group them conceptually and order them narratively.

The first two filters reflect the logic that a CAMEO user's CH often provides correct answers to the CM fairly quickly; thus, the distribution of $\Delta t_{m,h,c}$ over items $i$ should be positive with small magnitudes. Fig. 2 shows four contrasting distributions of $\Delta t_{m,h,c}$ for four different CH-CM pairs. Distribution A illustrates two unrelated and asynchronous accounts, where one user's "show answer" event is sometimes before and sometimes after another user's correct answer submissions by times that vary widely in magnitude; distributions like this should be common. Distribution B illustrates two users (e.g. siblings, roommates, or students taking the assessment side-by-side) working in close synchronicity. Due to chance and differences in pacing, one user's "show answers" will sometimes precede but sometimes follow the other's "correct answers," but times will be in close proximity.

Distribution C reflects prototypical CAMEO behavior, corresponding to Fig. 1. All $\Delta t_{m,h,c}$ are positive, and their magnitudes are extremely small, centered in this illustration at around 10 seconds. These small $\Delta t_{m,h,c}$ magnitudes are typically possible when the CAMEO user is logged in simultaneously to both CH and CM accounts on different internet browsers or computers. Finally, Distribution D is also positive but with $\Delta t_{m,h,c}$ magnitudes that are larger and more variable. This is consistent with ordered coincidence, where unrelated pairs of users will be offset from each other due to different enrollment dates or time-of-day preferences.

To identify CAMEO users by distributions of $\Delta t_{m,h,c}$, we considered constraining the population distribution of $\Delta t_{m,h,c}$ or $|\Delta t_{m,h,c}|$ by strong parametric assumptions (e.g., log-normal, exponential), but many observed distributions had extreme skew due to outlying $\Delta t_{m,h,c,i}$ values. We therefore opt for a less parametric approach that targets the percentage of positive observations (Filter 1) and the magnitude of the 90th percentile (Filter 2).

*Filter 1: The Bayesian criterion*

For Filter 1, given variation in the quantity of data shared between any CH and CM, we use a Bayesian criterion that is more stringent when data are limited (Lehmann & Casella, 1998). We estimate the parameters of the posterior distribution of a proportion $\pi$, our parameter of interest indicating the proportion of positive $\Delta t_{m,h,c,i}$ values, given $n$, as the number of in-common items for which a CH has a "show answer" and a CM has a correct answer, and $x$, as the number of times that the CH time precedes the CM time:

$$x_{m,h,c} = \sum_{i=1}^{n} I(\Delta t_{m,h,c,i} > 0)$$

Here, $I$ is the indicator function, which is 1 when the argument is true and 0 otherwise. The maximum $n$ for any CH-CM pair is the number of items. The average number of graded items is 141, across courses, allowing considerable data for inference. We assume that $x$ is binomially distributed and that $\pi$ has a Beta distribution. Following standard rules of conjugacy:

$$x|n, \pi \sim \text{Binomial}(\pi, n)$$
$$\pi|\alpha, \beta \sim \text{Beta}(\alpha, \beta)$$
$$\pi|x, n, \alpha, \beta \sim \text{Beta}(\alpha + x, \beta + n - x)$$

We observe $x$ and $n$ in the data. For the prior distribution, we set $\alpha = \beta = 0.5$, empirically and judgmentally, using full distributions of observed $p = x/n$ when $n$ is large in our data. This is a gentle U-shape, consistent with the fact that many distributions of $t_{m,c}$ are stochastically or entirely offset from other distributions of $t_{h,c}$ in one direction or other, due to the asynchronous nature of MOOCs.

We operationalize Filter 1 in terms of confidence that $\pi$ is close to 1, that is, that CH and CM are almost always ordered respectively. Specifically, Filter 1 selects CH-CM pairs with a 90% probability of $\pi_{m,h,c} > 0.9$. This is a conservative, stringent criterion that requires considerable data before concluding that a distribution is predominantly positive. Even a CH-CM pair with $x = 12$ out of $n = 12$ ($p = 100\%$) positive values is insufficient to meet this criterion.

*Filter 2: Setting the cutoff threshold*

Filter 2 addresses the fact that Filter 1 excludes Distributions A and B from CAMEO consideration, but it cannot distinguish between Distributions C and D (Fig. 2). To exclude ordered accounts that happen to be offset in time in the positive direction, Filter 2 uses the 90th percentile of the $\Delta t_{m,h,c}$ distribution as a criterion, setting a conservative cutoff at 5 minutes. In other words, 90% of the $\Delta t_{m,h,c}$ values must be less than 5 minutes. This cutoff occurs at an "elbow" as shown in Fig. 3, where shifting the cutoff between 0 and 5 minutes changes the number of estimated CAMEO users dramatically, and subsequent shifts past 5 minutes do not.

*Filter 3: Certified CM – uncertified CH pairs.*

The first two filters provide considerable evidence that, for CAMEO users, the distribution of $\Delta t_{m,h,c}$ is disproportionately positive and centered at less than 5 minutes in time. Filters 3 through 5 provide convergent criteria to further minimize the probability of false identification.

Filter 3 considers only CH-CM pairs for which the CH is uncertified and the CM is certified. Although this may discard CAMEO users who do not ultimately earn certification, our intention is to address possible threats to MOOC certificate validity as directly as possible, so we include only certified CMs. In addition, a CH that earns a certificate is inconsistent with the interpretation of CAMEO users as a cheating strategy, since it leaves open the possibility that the CH is actually proficient in the course.

*Filter 4: Detecting Shared IP address*

Filter 4 further reduces the candidate pool to those CH-CM pairs who share an IP address, defined for each account as the modal (most commonly used) IP address across all logged interactions in a given course $c$. However, considering only users with the same IP address fails to detect users who employ the CAMEO strategy using accounts assigned different modal IP addresses in a given course, either by coincidence or intentional misdirection. To improve detection of these users, we broaden the definition of "sharing an IP address" to CH-CM pairs who have ever shared an IP address in their course-taking history.

To detect CAMEO users with accounts having different modal IP addresses in a given course, we consider every unique (name, IP) tuple across all accounts participating in any of the 115 courses analyzed. We assign each (name, IP) an "IP group", initially as a unique integer for each pair. Next, we group by modal IP address such that all (name, IP) tuples sharing the same modal IP address are assigned (merged into) the same IP group. Then, we group by username such that all (name, IP) tuples sharing the same username are merged into the same IP group. We repeat both the "merge by IP" and "merge by username" steps until the IP group no longer changes. This can be described as a "transitive closure" of modal IP address and account names for all accounts across courses. It allows us to consider CM-CH pairs whenever the two accounts have shared a common modal IP address within a course, across courses, or across other accounts that have shared the same modal IP address within and across courses.

*Filter 5: Excluding shared routers*

Filter 5 excludes all CH-CM pairs who are part of a group that has 10 accounts or more that share a modal IP address. We intend this to exclude shared routers among classrooms or cafes that might increase the likelihood of false positives.

## RESULTS

We investigate[¥] the prevalence of CAMEO users in 115 online courses from two institutions, Harvard University and MIT, offered on the MOOC platform, edX. We use data from courses from the fall of 2012 through the spring of 2015, up to an analytic cutoff date of June 2, 2015. About half of these MOOCs are described in detail in other reports (McNutt, 2013; Ho et al., 2015) that emphasize their range of curricular foci and their heterogeneous participant demographics. Our sample consists of 1,893,092 enrollments (1,067,570 from unique accounts) whose users clicked into the course content at least once. A total of 155,301 certificates were ultimately earned from 103,370 unique accounts.

### Prevalence of CAMEO

Across these courses, we estimate that a total of 1,237 certificates were earned using the CAMEO strategy, 1% across all 115 courses, by 657 unique users employing 674 harvester accounts. In some courses, CAMEO users account for as many as 5% of the certificates earned. Across the 69 courses in which we identified CAMEO users, they account for 1.3% of certificates. Table 2A shows that CAMEO users are more likely to be young, male, less educated, and international than their certified counterparts in the same courses (Ho et al., 2015). Among countries with at least 20 CAMEO users, countries with the highest CAMEO counts per certificates were Albania (12%), Indonesia (4%), Serbia (3%), Colombia (2%), and China (2%). The CAMEO rate in the USA is particularly low, at 0.4% of certificates earned. Table 2B shows CAMEO prevalence by broad curricular area. Prevalence of CAMEO users is greatest in the Government, Health, and Social Science category

---

[¥] A list of the 115 courses studied, with their classifications into topic areas, and $\Delta t_{m,h,c}$ distribution data for CM-CH pairs, are archived in the Harvard Dataverse Network, at http://dx.doi.org/10.7910/DVN/3UKVOR.

(1.3%) and lowest in the Computer Science category (0.1%).

### *Prevention of CAMEO*

Mechanisms which logically prevent CAMEO use include restricting the "show answer" option until after assignments are due, and using algorithmic generation of assessment items so that participants receive randomly varying items, each with different solutions. Across the 37 Science, Technology, Engineering, and Mathematics (STEM) courses in this sample, 18 employed such prevention mechanisms. Table 2C shows that the CAMEO rate in courses that employed these preventive strategies in half or more of the assessment items was substantially lower (0.1%) than the rate in courses that did not employ preventive strategies (1.2%).

### DISCUSSION

As open online courses proliferate, we identify CAMEO as a significant threat to the validity of large-scale certification. Our primary goals are to demonstrate that CAMEO exists and to bound its prevalence in the population. We believe that our method accomplishes this and does so conservatively. Nonetheless, we raise here a central shortcoming of this work and address it briefly while encouraging subsequent research. Like many cheating analyses in real contexts, we have no "true" knowledge of cheating to evaluate whether our detection method is accurate at the individual level. Perhaps a child is guessing haphazardly and clicking "show answer," while working with a parent who separately submits answers correctly, always a few minutes after the child. This is unlikely but not impossible.

We raise briefly here three convergent sources of evidence. First, text-matching of usernames reveals considerable overlap in candidate pairs; many CAMEO users have usernames consistent with the Master-Harvester hypothesis, like "Curtis1" and "Curtis2." Second, although our CAMEO detection algorithm treats every CM-CH pair independently, we find CAMEO behavior is clustered within users. A total of 43 separate accounts have earned 5 or more certificates by CAMEO. Third, we conducted a limited analysis, in one course, of plagiarism by copying open-response text across users, and we find that these accounts are also identified as CAMEO users. Although we believe our algorithm alone is sufficient to demonstrate the existence and bound the prevalence of CAMEO, we encourage further research to support validation of the detection algorithm.

Another concern is the possibility that some users could be using CAMEO to increase their exposure to assessment items and thereby increase their learning. We argue that this is unlikely given how we operationalize our definition. CAMEO users require nearly all of CH "show answer" clicks to occur "shortly" before CM correct answer submissions. In fact, we found that often the actual time difference was only a few a seconds. The extent and timing of this systematic behavior is most consistent with a cynical and blatant attempt to harvest correct answers to rapidly acquire certification, not with a learning strategy.

Our estimates of cheating prevalence are arguably consistent with higher estimates from surveys. Such surveys typically ask a variant of the question "Have you cheated?" with allowance for recency and magnitude (McCabe, Butterfield, & Treviño, 2012). In contrast, CAMEO is complete in its scope and course-specific, as the introduction notes. The analogous question we address is, "Did you cheat your way through this entire course?" We can establish a basis for comparison through the observation from our data, that those who certify in multiple courses are much more likely to have used the CAMEO strategy at least once, including 25% of those who have earned at least 20 certificates, as depicted in Table 3. We consider this commensurate in severity to the reports that two-thirds of college students have engaged in some form of academic dishonesty in the previous year (McCabe, Butterfield, & Treviño, 2012), especially considering that the minimum threshold in our analysis is sufficient cheating to earn certification, versus being dishonest in just one or a few problems.

Our findings are consistent with other observations that MOOC assessment infrastructures rarely support robust inferences about learning (Reich, 2015). All feasible mechanisms that prevent the CAMEO strategy have a downside. If instructors withhold the "show answer" option until after the problems are graded, this would constrain generally desirable asynchronous MOOC usage, and students will not have the rapid feedback touted as a pedagogical benefit of online learning environments. Algorithmic generation of assessment items and correct answers is challenging and only suitable for some subjects and assessment tasks.

Beyond honor codes (LoSchiavo & Shatz, 2011; Corrigan-Gibbs, Gupta, Northcutt, Cutrell, & Thies, 2015), a solution embraced by many MOOC purveyors (Kolowich, 2013; Straumsheim, 2015; Eisenberg, 2013) is to offer certificates earned under controlled assessment conditions, such as in-person assessments taken at secure testing centers for a fee. We observe that the cost and constraints associated with fee-based, in-person testing centers are antithetical to the open, online principles that define MOOCs, as well as their mission of improving worldwide access to not just learning but certification opportunities. Further research on cheating detection and prevention, including experiments that can isolate factors that cause and discourage cheating, is necessary to design spaces and structures that can support open and trustworthy certification at scale.

## CONCLUSION

The CAMEO detection algorithm uses three strategies that hold general promise for the analysis of clickstream data. First, time difference analysis is a tool to infer relationships among students. Second, Bayesian criteria allow appropriately conservative classification when data are limited. Third, transitive closure is a technique for robust consideration of possible CAMEO users. Beyond cheating detection in MOOCs, these tools may aid more generally in identification of collaboration and interaction among online users.

There is continued interest in the potential for MOOCs to increase efficiency and spur innovation in higher education. Four features of CAMEO severely undermine this potential. First, unless prevented, this cheating strategy allows students to earn certificates in open online courses without any understanding of the domain material. Second, the strategy is highly convenient, requiring no interactions with external resources, either animate or inanimate. Third, it is unrestricted, employable in a nonselective, open admission setting. Fourth, whereas cheating is traditionally considered with respect to individual assessments or portions thereof, CAMEO is a course-level strategy. It is less cheating than the wholesale falsification of a certificate.

In this paper, we have demonstrated the prevalence of the CAMEO cheating strategy, and we have argued that it poses a serious threat to interpretations of MOOC certification. Protecting certification requires CAMEO prevention, and we have shown that preventive strategies hold promise. Yet, CAMEO is only one of many possible cheating strategies. Sophisticated detection algorithms should be a part of a general approach to protect the validity of online course certification. We recommend and look forward to future interventions that increase and encourage honest behavior in online learning environments while disallowing and discouraging cheating in all its forms.

## ACKNOWLEDGEMENTS

This material is based upon work supported by the National Science Foundation Graduate Research Fellowship under Grant No. (#1122374). We express our thanks to the HarvardX and MITx staff for their continued support, and in particular, we are grateful to the ChinaX team for their input in the early stages of this research.

## REFERENCES

Alraimi, K. M., Zo, H., & Ciganek, A. P. (2015). Understanding the MOOCs continuance: The role of openness and reputation. *Computers & Education*, *80*, 28-38.

Baker, R. S., Corbett, A. T., & Koedinger, K. R. (2004, January). Detecting student misuse of intelligent tutoring systems. In *Intelligent tutoring systems* (pp. 531-540). Springer Berlin Heidelberg.

Christensen, G., Steinmetz, A., Alcorn, B., Bennett, A., Woods, D., & Emanuel, E. J. (2013). The MOOC phenomenon: who takes massive open online courses and why? http://dx.doi.org/10.2139/ssrn.2350964

Corrigan-Gibbs, H., Gupta, N., Northcutt, C., Cutrell, E., & Thies, W. (2015, March). Measuring and Maximizing the Effectiveness of Honor Codes in Online Courses. In *Proceedings of the Second (2015) ACM Conference on Learning@ Scale* (pp. 223-228). ACM. http://dx.doi.org/10.1145/2724660.2728663

Coursera. (2012, April). Terms of Use, Privacy Policy and Honor Code. Retrieved from https://authentication.coursera.org/auth/auth/normal/tos.php

DeBoer, J., Ho, A. D., Stump, G. S., & Breslow, L. (2014). Changing "Course" Reconceptualizing Educational Variables for Massive Open Online Courses. *Educational Researcher*. http://dx.doi.org/10.3102/0013189X14523038

d'Oliveira, C., Carson, S., James, K., & Lazarus, J. (2010). MIT OpenCourseWare: Unlocking knowledge, empowering minds. *Science*, *329*(5991), 525-526.

edX. (2014, October 22). edX Terms of Service. Retrieved from https://www.edx.org/edx-terms-service

edX. (2015). Student FAQ. Retrieved from https://www.edx.org/about/student-faq

Eisenberg, A. (2013, March 2). Keeping an eye on online test-takers. *The New York Times*. Retrieved from http://www.nytimes.com/2013/03/03/technology/new-technologies-aim-to-foil-online-course-cheating.html

Ho, A. D., Reich, J., Nesterko, S. O., Seaton, D. T., Mullaney, T., Waldo, J., & Chuang, I. (2014). HarvardX and MITx: The first year of open online courses, Fall 2012-Summer 2013. http://dx.doi.org/10.2139/ssrn.2381263

Ho, A. D., Chuang, I., Reich, J., Coleman, C. A., Whitehill, J., Northcutt, C. G., ... & Petersen, R. (2015). HarvardX and MITx: Two Years of Open Online Courses Fall 2012-Summer 2014. http://dx.doi.org/10.2139/ssrn.2586847

Hoxby, C. M. (2014). *The economics of online postsecondary education: MOOCs, nonselective education, and highly selective education* (No. w19816). National Bureau of Economic Research.

Kafai, Y. B., & Fields, D. A. (2009). Cheating in virtual worlds: Transgressive designs for learning. *On the horizon*, *17* (1), 12-20. http://dx.doi.org/10.1108/10748120910936117

Kauffman, Y., & Young, M. F. (2015). Digital plagiarism: An experimental study of the effect of instructional goals and copy-and-paste affordance. *Computers & Education*, *83*, 44-56.

Kolowich, S. (2013, April 15). Behind the Webcam's Watchful Eye, Online Proctoring Takes Hold. *The Chronicle of Higher Education*. Retrieved from http://chronicle.com/article/Behind-the-Webcams-Watchful/138505/

Lehmann, E. L., & Casella, G. (1998). *Theory of point estimation* (Vol. 31). Springer Science & Business Media.

Linn, M. C., Gerard, L., Ryoo, K., McElhaney, K., Liu, O. L., & Rafferty, A. N. (2014). Computer-guided inquiry to improve science learning. *Science*, *344* (6180), 155-156.

LoSchiavo, F. M., & Shatz, M. A. (2011). The impact of an honor code on cheating in online courses. *MERLOT Journal of Online Learning and Teaching*, *7*(2).

Mastin, D. F., Peszka, J., & Lilly, D. R. (2009). Online academic integrity. *Teaching of Psychology*, *36* (3), 174-178.

McCabe, D. L., Butterfield, K. D., & Trevino, L. K. (2012). *Cheating in college: Why students do it and what educators can do about it*. JHU Press.

McNutt, M. (2013). Bricks and MOOCs. *Science*, *342* (6157), 402-402.

Palazzo, D. J., Lee, Y. J., Warnakulasooriya, R., & Pritchard, D. E. (2010). Patterns, correlates, and reduction of homework copying. *Physical Review Special Topics-Physics Education Research*, *6*(1), 010104.

Pappano, L. (2012). The Year of the MOOC. *The New York Times*, *2* (12). Retrieved from http://www.nytimes.com/2012/11/04/education/edlife/massive-open-online-courses-are-multiplying-at-a-rapid-pace.html

Reich, J. (2015). Rebooting MOOC research. *Science*, *347* (6217), 34-35.

Smith, M. S. (2009). Opening education. *Science*, *323* (5910), 89-93.

Stanford Online (2013). "Harnessing new technologies and methods to advance teaching and learning at Stanford and beyond." *Stanford Online 2013 in Review*.

Straumsheim, C. (2015, April 23). MOOCs for (a Year's) Credit. *Inside Higher ED*. Retrieved from https://www.insidehighered.com/news/2015/04/23/arizona-state-edx-team-offer-freshman-year-online-through-moocs

Udacity. (2014, June 28). Udacity Terms of Service. Retrieved from https://www.udacity.com/legal/tos

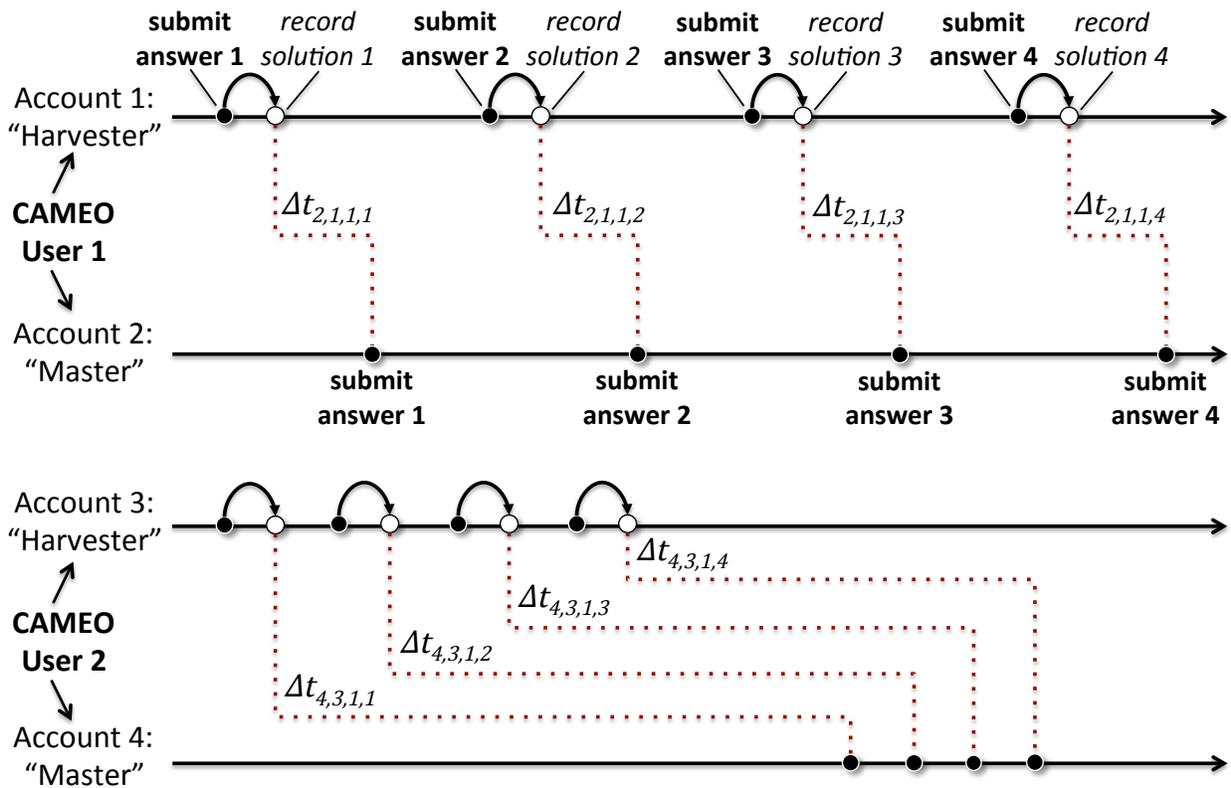

**Fig 1.** Two types of prototypical behavior when Copying Answers using Multiple Existences Online (CAMEO). A "harvester" account $h$ records correct solutions, and a "master" account $m$ submits correct answers. The time between harvesting in account $h$ (white dot) and correct answer submission by account $m$ (black dot) is estimable from the data and defined as $\Delta t_{m,h,c,i}$ for item $i$ in course $c$. The strategy employed by CAMEO 1 is to alternate harvesting and submission. The strategy of CAMEO 2 is to harvest a cluster and then submit a cluster.

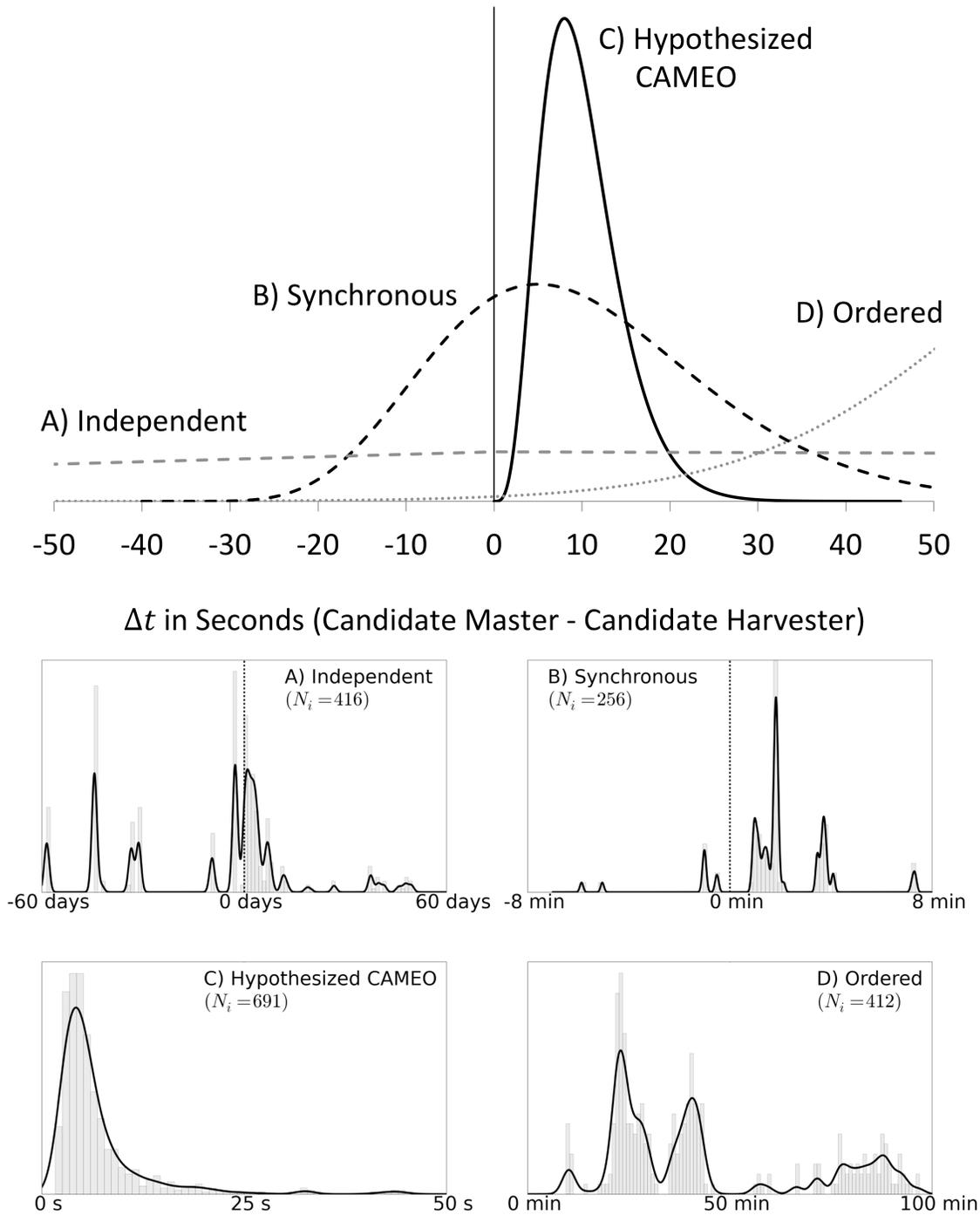

**Fig 2.** Four theoretical distributions of Δ*t* (top) illustrated by empirical distributions (below). Distribution A illustrates uniformly distributed "show answer" and "correct submission" times resulting in a shallow triangular distribution symmetrical around 0. Distribution B illustrates synchronous submission with positive and negative Δ*t* values. Distribution C illustrates prototypical "Copying Answers using Multiple Existences Online" (CAMEO) behavior, with candidate harvester accounts passing solutions to candidate master accounts over a short time span. Distribution D illustrates consistently and coincidentally ordered submissions over a longer time span. For the empirical distributions, the number of items shared between a harvester's "show answer" and a master's "correct submission" is displayed as $N_i$.

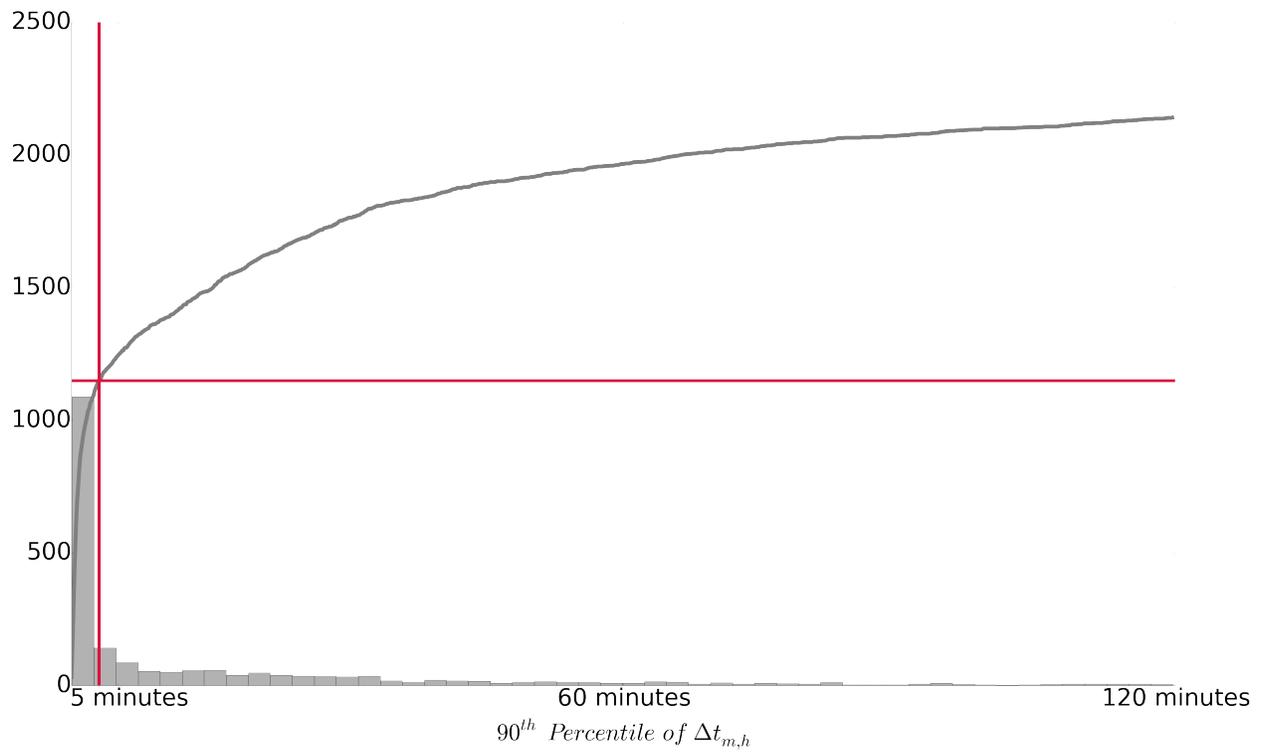

**Fig. 3.** Cumulative distribution (line) showing number of CAMEO users identified versus the 90$^{th}$ percentile cutoff value of $\Delta t_{m,h}$ (Filter 2 in Table 1), together with the associated histogram (bars). The vertical red line depicts the cutoff value chosen; the horizontal red line is the corresponding number of CAMEO users identified.

**Table 1.** A detection approach that asserts five necessary filtering conditions for candidate harvester (CH) and candidate master (CM) pairs to be classified as Copying Answers using Multiple Existences Online (CAMEO).

| Condition | Explanation | Operationalization |
|---|---|---|
| 1) The $\Delta t$ distribution should be positive | The CH should harvest the correct answer before the CM submits the correct answer. | Bayesian – 90% confident that the proportion of positive $\Delta t$ values is 90%. |
| 2) The magnitudes of $\Delta t$ should be small | The CH should provide answers to the CM quickly. | The $90^{\text{th}}$ percentile of the $\Delta t$ distribution should be less than 5 minutes. |
| 3) The CH should not be certified, and the CM should be certified | The CH should be guessing and uninterested in certification, whereas the goal of the CM is presumably certification. | A CM must be certified. A CH must not be certified. |
| 4) The CM and CH should share an IP address or have shared one at some point in their course-taking history. | This increases the likelihood that the CM and CH are in fact the same person. | The CM and CH must share one of the sets determined by the transitive closure of modal IP address and account name over courses. |
| 5) There should be few accounts that share or have shared an IP address with the CM and CH. | This excludes internet cafes, school networks, and other common spaces where chance coincidence of $\Delta t$ may lead to false detection. | The number of accounts with a shared modal IP address must not exceed 10. |

*Notes*. The filters are chosen to be conservative, and their conjunctive application is more so, minimizing the chance of false identification at the cost of conceding missed CAMEO users. In terms of missed identification, Filter 1 excludes small-sample CAMEO users even when their proportions of positive times are 100%. Filter 2 excludes CAMEO users that take more than 5 minutes to pass solutions between accounts. Filter 3 excludes those who use the CAMEO strategy but do not earn certificates. Filter 4 addresses those who use IP-masking strategies like the Tor browser. Filter 5 excludes CAMEO users within classrooms, cafes, and other scenarios in which IP addresses are shared.

**Table 2.** Distribution and demographics of those identified as Copying Answers using Multiple Existences Online (CAMEO users) across courses. (A) Prevalence and demographic distribution of CAMEO users versus non-CAMEO certificate earners in the 69 courses with nonzero CAMEO users. (B) Distribution of CAMEO users across four broad curricular areas, for the 115 courses in the dataset. (C) Observed differences in CAMEO percentage for Science, Technology, Engineering, and Math courses that do or do not employ mechanisms that logically prevent CAMEO users, including solutions embargoed until after due dates and algorithmic generation of problems with varying solutions.

(A)

| Among 69 Courses with CAMEO users | Non-CAMEO | CAMEO |
|---|---|---|
| N Certified | 96,367 | 1,237 (1.3%) |
| % Female | 33% | 19% |
| % Bachelor's | 79% | 59% |
| Median Age | 32 | 25 |
| % USA | 30% | 14% |

(B)

| Among All 115 courses | N Courses | % CAMEO of Certified |
|---|---|---|
| Computer Science | 12 | 0.1% |
| Government, Health, and Social Science | 28 | 1.3% |
| Humanities, History, Religion, Design, and Education | 38 | 1.1% |
| Science, Technology, Engineering, and Mathematics | 37 | 0.7% |
| Overall | 115 | 0.9% |

(C)

| Among 37 STEM courses | N Courses | N Certified | N CAMEO | % CAMEO (typical user) | % CAMEO (typical course) |
|---|---|---|---|---|---|
| No or limited CAMEO prevention | 19 | 19,383 | 171 | 0.9% | 1.2% |
| CAMEO prevention | 18 | 11,717 | 8 | 0.1% | 0.1% |
| Overall | 37 | 31,100 | 179 | 0.6% | 0.7% |

*Note*: Survey methods follow those of other studies: Demographic information collected from edX surveys with response rates >95%; Country is determined by geolocation of the modal IP address; Courses are divided into curricular areas judgmentally.

**Table 3.** Rates of Copying Answers using Multiple Certificates Online (CAMEO) among unique accounts earning multiple certificates.

| Number of Certificates: $N$ (Lower Bound) | Unique Certificate Earners with $\geq N$ Certificates: $M$ | Unique Certificate Earners, $M$, with $\geq 1$ CAMEO | Percent of Unique Certificate Earners with $\geq 1$ CAMEO |
|---|---|---|---|
| 1 | 103,370 | 657 | 1% |
| 5 | 3,435 | 185 | 5% |
| 10 | 1,262 | 82 | 6% |
| 15 | 200 | 35 | 18% |
| 20 | 73 | 18 | 25% |
| 25 | 35 | 14 | 40% |
| 30 | 15 | 7 | 47% |
| 40 | 3 | 2 | 67% |